\documentclass[12pt]{article}
\usepackage[dvips]{epsfig}
\begin{document}
\begin{center}
\large{\bf A Summary of Charm Hadron Production Fractions} \\
\normalsize
\vspace{0.5cm}
Erich Lohrmann, University of Hamburg \\ 
\end{center}

Charm hadron production fractions are given by the probability that a c-quark
fragments into a given hadron. \\

Charm hadron production fractions have been measured at HERA by the H1
Collaboration [1] in DIS, and by the ZEUS Collaboration in DIS [2],[3]
and in photoproduction [4]. In addition there is a summary of measurements
at $e^+e^-$ storage rings [5], updated in ref.[4].\\
In the meantime new decay branching ratios for charmed hadrons have been 
published. The charm hadron production
fractions have therefore to be updated, using the results from the 2010
PDG [6] on branching ratios. \\

The new relevant branching ratios are [6] : \\
\noindent
$D^+ \to K^{-}2\pi^+=0.094 \pm 0.004$ \\
$D^0 \to K^{-}\pi^+ =0.0389 \pm 0.0005$ \\
$D^* \to D^0 \pi^+ =0.677 \pm 0.005 $ \\
$D_s \to \Phi(\to K^+K^{-})\pi = 0.0232 \pm 0.0014$ \\
$\Lambda_c \to K p \pi = 0.050 \pm  0.013 $ \\ 

 This note uses updated values produced by L. Gladilin [7]
 for e+e- storage rings from the LEP, ARGUS and CLEO Collaborations, and for 
HERA photoproduction data.
DIS data from  HERA from the H1 and ZEUS Collaborations,
 also updated to 2010 branching ratios, are added here. \\

Tables 1 and 2 show the updated input for the charm hadron production 
fractions. \\

\begin{table}[h]%
\centering
\caption{Input}
\begin{tabular}{|r|r r|r r|}
\hline
   &     ZEUS $\gamma$ p &         &     ZEUS DIS &                \\
   &        stat.      & syst.      &    stat.        & syst.     \\
\hline
$D^+$& 0.222$\pm$0.015 &+0.014-0.005& 0.217$\pm$0.018 & +0.002 -0.019 \\
$D^0$& 0.532$\pm$0.022 &+0.018-0.017& 0.585$\pm$0.019 & +0.009 -0.052 \\
$D_s$& 0.075$\pm$0.007 &+0.004-0.004& 0.086$\pm$0.010 & +0.007 -0.008 \\
$\Lambda_c$&0.150$\pm$0.023&+0.014-0.022&0.098$\pm$0.027& +0.020 -0.017   \\
$D^*$& 0.203$\pm$0.009 &+0.008-0.007& 0.234$\pm$0.011 & +0.006 -0.021 \\
\hline
\end{tabular}
\end{table}

\newpage

\begin{table}[h]%
\centering
\caption{Input}
\begin{tabular}{|r|r|r|}
\hline
   &                 e+e-      & H1 DIS \\
   &               stat.+syst  & stat.+syst.\\
\hline
$D^+$&  0.222$\pm$0.010&0.204$\pm$0.026\\
$D^0$&  0.544$\pm$0.022&0.584$\pm$0.048\\
$D_s$&  0.077$\pm$0.006& 0.121$\pm$0.044\\
$\Lambda_c$& 0.076$\pm$0.007 &   \\
$D^*$&  0.235$\pm$0.007& 0.276$\pm$0.034\\
\hline
\end{tabular}
\end{table}

The sum of fractions for the HERA data is normalised by  imposing the
 condition \\
$f(D^+)+f(D^0)+f(D_c)+1.14 \cdot f(\Lambda_c)=1.0$. \\

For the $e^+e^{-}$ data the total charm production cross-section is known in
principle from QCD. Therefore the charm fractions should as a check add
automatically to 1.0 (with a factor 1.14 for $f(\Lambda_c)$ to account for
unobserved baryon states).
 The sum in table 2 is 0.93, differing from 1.0 by about
2 s.d.. Table 3 shows the $e^{+}e^{-}$ data normalized to 1.0, taking into
account statistical, systematic and branching ratio uncertainties, added in
quadrature. 
The charm fraction for the $D^*$ is not changed in this normalisation process.
This rests on the assumption, that the total charm cross-section given by QCD
is appropriate for normalisation. \\


\begin{table}[h]%
\centering
\caption{Input $e^+e{-} $ data, normalized to 1.0}
\begin{tabular}{|r|r|}
\hline
      &  stat.+syst. \\
\hline
$D^+$ &0.232$\pm$0.010  \\
$D^0$ & 0.573$\pm$0.023  \\
$D_s$ & 0.080$\pm$0.006  \\
$\Lambda_c$ &  0.104$\pm$0.010 \\
$D^*$ &  0.235$\pm$0.007  \\
\hline
\end{tabular}
\end{table}

The uncertainties due to the 2010 branching ratio uncertainties
 are shown in Table 4.

\begin{table}[h]%
\centering
\caption{Uncertainties of charm fractions due to branching ratio uncertainties}
\begin{tabular}{|r|r r|r r|r r|r r|}
\hline
   & ZEUS $\gamma$ p &   & ZEUS DIS     &   & e+e-         &  & H1 DIS       &      \\
\hline
$D^+$&  +0.011& -0.013    & +0.009 & -0.010    & +0.010 & -0.009 & +0.009 & -0.010 \\
$D^0$&  +0.019& -0.028    & +0.018 & -0.019    & +0.007 & -0.007 & +0.018 & -0.019 \\
$D_s$&  +0.005& -0.005    & +0.005 & -0.005    & +0.005 & -0.004 & +0.008 & -0.008   \\
$\Lambda_c$&+0.038& -0.025& +0.025 & -0.023    & +0.027 & -0.016 &        &       \\
$D^*$&  +0.007& -0.010    & +0.007 & -0.010    & +0.003 & -0.003 & +0.009 & -0.012  \\
\hline
\end{tabular}
\end{table}

The numbers of these experimental inputs (with the $e^+e^{-}$ data normalized
 to 1.0) are now combined. Statistical
and systematic uncertainties are added in quadrature for each experiment. \\ 

Taking  asymmetric systematic uncertainties into account has a negligeable
 effect.
Uncertainties from the branching ratios, which are correlated, are ignored.\\

Table 5 shows the  result of the averaging process. 
The uncertainties are statistical and systematic
added in quadrature.
The branching ratio uncertainties are weighted averages of the uncertainties
given in table 4.

\begin{table}[h]%
\centering
\caption{ Mean c-fractions}
\begin{tabular}{|l|l|l|l l|}
\hline
          &                    & av. pull & br.ratio unc.& \\
\hline
$D^+$    &    0.2256 $\pm$ 0.0077 & 0.57 & +0.010  & -0.010  \\
$D^0$    &    0.5643 $\pm$ 0.0151 & 0.70 & +0.0135 &  -0.0164 \\
$D_s$    &    0.0797 $\pm$ 0.0045 & 0.61 & +0.0052 &  -0.0046 \\
$\Lambda_c$ &  0.1080 $\pm$ 0.0091 &0.88 & +0.0279 &  -0.0174 \\
$D^*$    &    0.2287 $\pm$ 0.0056 & 1.38 & +0.0045 &  -0.0056 \\
\hline
\end{tabular}
\end{table}

The sum of the fractions is 0.993, the difference to 1.0 is well within the 
uncertainties.

\newpage

{\bf References:}\\

\noindent
(1) H1 Collab.,A.Aktas et al.,Eur.Phys.J.C38(2005)447 \\
(2) ZEUS Collab.,S.Chekanov et al.,JHEP 07(2007)74\\
(3) ZEUS Collab.,H.Abramowicz et al., JHEP11(2010)009 \\
(4) ZEUS Collab.,S.Chekanov et al.,Eur.Phys.J.C44(2005)351 \\
(5) L.Gladilin, Preprint hep-ex/9912064,1999 \\
(6) Particle Data Group ,J.of Physics G, Vol.37 No 7A (2010) \\
(7) ATLAS Collaboration, ATLAS-CONF-2011-017,2011 \\
\end{document}